\documentstyle[preprint,prl,aps]{revtex}

\begin{document}
\preprint{\vbox{\hbox{July 1994}\hbox{IFP-702-UNC}\hbox{VAND-TH-94-8}}}
\draft
\title{Simple Non-Abelian Finite Flavor Groups and Fermion Masses}
\author{\bf P. H. Frampton$^{(a)}$ and T. W. Kephart$^{(a,b)}$}
\address{(a)Institute of Field Physics, Department of Physics and Astronomy,\\
University of North Carolina, Chapel Hill, NC  27599-3255}
\address{(b)Department of Physics and Astronomy,Vanderbilt University,
Nashville, TN 37235\footnote{Permanent address}}
\maketitle

\begin{abstract}
The use of nonabelian discrete groups G as family symmetries is discussed
in detail. Out of all such groups up to order g = 31, the most
appealing candidates are two subgroups of SU(2): the dicyclic [double
dihedral] group G = $Q_6 ={ }^{(d)}D_3$ ( g = 12 ) and the double
tetrahedral group $^{(d)}T = Q_4\tilde{\times}Z_3$ ( g = 24 ).
Both can allow a hierarchy $t > b, \tau > c > s, \mu > u, d, e$.
The top quark is uniquely allowed to have a G symmetric mass. Sequential
breaking of G and radiative corrections give the smaller masses.
Anomaly freedom for gauging $G \subset SU(2)$ is a strong constraint
in assignment of fermions to representations of G.
\end{abstract}
\newpage

{\bf I Introduction}\\

To cite just a few of very many examples, discrete groups are central
to molecular orbitals just as the crystallographic groups are in
solid-state physics; the discrete symmetries C, P and T and their
violation have made a profound impact on our understanding of quantum
field theory. Thus, finite groups play a major role in physics.

In the minimal standard model, nine of the nineteen parameters
are quark and lepton masses, and four more pertain to quark flavor mixing.
Thus more than 2/3 of the parameters are associated with fermion masses.
The fermion masses range from the electron mass (0.511MeV) up to the
top quark mass (174GeV), a ratio of 340,000. With a single Higgs
doublet, this implies a corresponding puzzling hierarchy in the Yukawa coupling
constants.

This hierarchy can be ameliorated by the postulation of a flavor or
family symmetry group G underlying the mass pattern. The G we employ will
be nonabelian because a nonabelian group has nontrivial multidimensional
representations; G will also be taken as finite because this allows
more low dimensional representations and hence more structure than
possible for a continuous Lie symmetry group.
Nevertheless, it will be important that G be gaugeable to ensure consistency
with gravity, and this will constrain the assignments of fermions to
representations of G.

The possible finite groups up to order g = 31 ( It is traditional to stop
one below a power of 2 because of the proliferation of finite groups
for each g = $2^n$ ) are less than one hundred in number. They are all listed
in Section II below. The gauging of G is described in Section III.

Generalities of model building are in Section IV; specific examples,
most notably $Q_6$ and ${ }^{(d)}T$ are considered in Section V.
Finally, in Section VI are some concluding remarks.

\newpage
{\bf II Non-Abelian Finite Groups Up To Order 31}\\

{}From any good textbook on finite groups\cite{books} we may find a tabulation
of
the number of finite groups as a function of the order g, the number of
elements in the group. Up to order 31 there is a total of 93 different
finite groups of which slightly over one half (48) are abelian.

Amongst finite groups, the non-abelian examples have the advantage
of non-singlet irreducible representations which can be used to inter-relate
families. Which such group to select is based on simplicity: the minimum
order and most economical use of representations\cite{guts}.

Let us first dispense with the abelian groups. These are all made up from
the basic unit $Z_p$, the order p group formed from the $p^{th}$ roots
of unity. It is important to note that the the product $Z_pZ_q$ is identical
to $Z_{pq}$ if and only if p and q have no common prime factor.

If we write the prime factorization of g as:
\begin{equation}
g = \prod_{i}p_i^{k_i}
\end{equation}
where the product is over primes, it follows that the number
$N_a(g)$ of inequivalent abelian groups of order g is given by:
\begin{equation}
N_a(g) = \prod_{k_i}P(k_i)
\end{equation}
where $P(x)$ is the number of unordered partitions of $x$.
For example, for order $g = 144 = 2^43^2$ the value would be
$N_a(144) = P(4)P(2) = 5\times2 = 10$. For $g\leq31$ it is simple
to evaluate $N_a(g)$ by inspection. $N_a(g) = 1$ unless g contains
a nontrivial power ($k_i\geq2$) of a prime. These exceptions are:
$N_a(g = 4,9,12,18,20,25,28) = 2; N_a(8,24,27) = 3$; and $N_a(16) = 5$.
This confirms that:
\begin{equation}
\sum_{g = 1}^{31}N_a(g) = 48
\end{equation}
We shall not consider these abelian cases further in this paper.\\

Of the nonabelian finite groups, the best known are perhaps the
permutation groups $S_N$ (with $N \geq 3$) of order $N!$
The smallest non-abelian finite group is $S_3$ ($\equiv D_3$),
the symmetry of an equilateral triangle with respect to all
rotations in a three dimensional sense. This group initiates two
infinite series, the $S_N$ and the $D_N$. Both have elementary
geometrical significance since the symmetric permutation group
$S_N$ is the symmetry of the N-plex in N dimensions while the dihedral group
$D_N$ is the symmetry of the planar N-agon in 3 dimensions.
As a family symmetry, the $S_N$ series becomes uninteresting rapidly
as the order and the dimensions of the representions increase. Only $S_3$
and $S_4$ are of any interest as symmetries associated with the particle
spectrum\cite{Pak}, also the order (number of elements) of the $S_N$ groups
grow factorially with N. The order of the dihedral groups increase only
linearly with N and their irreducible representations are all one- and
two- dimensional. This is reminiscent of the representations of the
electroweak $SU(2)_L$ used in Nature.

Each $D_N$ is a subgroup of $O(3)$ and has a counterpart double dihedral
group $Q_{2N}$, of order $4N$, which is a subgroup of the double covering
$SU(2)$ of $O(3)$.

With only the use of $D_N$, $Q_{2N}$, $S_N$ and the tetrahedral group T ( of
order
12, the even permutations subgroup of $S_4$ ) we find 32 of the 45
nonabelian groups up to order 31, either as simple groups or as
products of simple nonabelian groups with abelian groups:
(Note that $D_6 \simeq Z_2 \times D_3, D_{10} \simeq Z_2 \times D_5$ and $
D_{14} \simeq Z_2 \times D_7$ )

$$\begin{tabular}{||c||c||}   \hline
g & \\    \hline
$6$  & $D_3 \equiv S_3$\\  \hline
$8$ & $ D_4 , Q = Q_4 $\\    \hline
$10$& $D_5$\\   \hline
$12$&  $D_6, Q_6, T$ \\ \hline
$14$& $D_7$\\  \hline
$16$& $D_8, Q_8, Z_2 \times D_4, Z_2 \times Q$\\  \hline
$18$& $D_9, Z_3 \times D_3$\\  \hline
$20$& $D_{10}, Q_{10}$ \\  \hline
$22$& $D_{11}$\\  \hline
$24$& $D_{12}, Q_{12}, Z_2 \times D_6, Z_2 \times Q_6, Z_2 \times T$,\\  \hline
 & $Z_3 \times D_4, D_3 \times Q, Z_4 \times D_3, S_4$\\  \hline
$26$& $D_{13}$\\  \hline
$28$& $D_{14}, Q_{14}$ \\  \hline
$30$& $D_{15}, D_5 \times Z_3, D_3 \times Z_5$\\  \hline
\end{tabular}$$
There remain thirteen others formed by twisted products of abelian factors.
Only certain such twistings are permissable, namely (completing all $g \leq 31$
)

$$\begin{tabular}{||c||c||}   \hline
g & \\    \hline
$16$  & $Z_2 \tilde{\times} Z_8$ (two, excluding $D_8$), $Z_4 \tilde{\times}
Z_4, Z_2 \tilde{\times}(Z_2 \times Z_4)$
(two)\\  \hline
$18$ & $Z_2 \tilde{\times} (Z_3 \times Z_3)$\\    \hline
$20$&  $Z_4 \tilde{\times} Z_7$ \\   \hline
$21$&  $Z_3 \tilde{\times} Z_7$ \\    \hline
$24$&  $Z_3 \tilde{\times} Q, Z_3 \tilde{\times} Z_8, Z_3 \tilde{\times} D_4$
\\  \hline
$27$&  $ Z_9 \tilde{\times} Z_3, Z_3 \tilde{\times} (Z_3 \times Z_3)$ \\
\hline
\end{tabular}$$

It can be shown that these thirteen exhaust the classification of {\it all}
inequivalent finite groups up to order thirty-one\cite{books}.

Of the 45 nonabelian groups, the dihedrals ($D_N$) and double dihedrals
($Q_{2N}$), of order 2N and 4N respectively,
form the simplest sequences. In particular, they fall into subgroups of
$O(3)$ and $SU(2)$ respectively,
the two simplest nonabelian continuous groups.

For the $D_N$ and $Q_{2N}$, the multiplication tables, as derivable from the
character tables,
are simple to express in general. $D_N$, for odd N, has two singlet
representations $1,1^{'}$ and $m = (N-1)/2$
doublets $2_{(j)}$ ($1 \leq j \leq m$). The multiplication rules are:

\begin{equation}
1^{'}\times 1^{'} = 1 ; ~~~1^{'}\times 2_{(j)} = 2_{(j)}
\end{equation}
\begin{equation}
2_{(i)}\times 2_{(j)} = \delta_{ij} (1 + 1^{'}) + 2_{(min[i+j,N-i-j])}
+ (1 - \delta_{ij}) 2_{(|i - j|)}
\end{equation}
\noindent

For even N, $D_N$ has four singlets $1, 1^{'},1^{''},1^{'''}$ and $(m - 1)$
doublets
$2_{(j)}$ ($ 1 \leq j \leq m - 1$)where $m = N/2$ with multiplication rules:

\begin{equation}
1^{'}\times 1^{'} = 1^{''} \times 1^{''} = 1^{'''} \times 1^{'''} = 1
\end{equation}
\begin{equation}
1^{'} \times 1^{''} = 1^{'''}; 1^{''} \times 1^{'''} = 1^{'}; 1^{'''} \times
1^{'} = 1^{''}
\end{equation}
\begin{equation}
1^{'}\times 2_{(j)} = 2_{(j)}
\end{equation}
\begin{equation}
1^{''}\times 2_{(j)} = 1^{'''} \times 2_{(j)} = 2_{(m-j)}
\end{equation}
\begin{equation}
2_{(j)} \times 2_{(k)} = 2_{|j-k|} + 2_{(min[j+k,N-j-k])}
\end{equation}

\noindent
(if $k \neq j, (m - j)$)

\begin{equation}
2_{(j)} \times 2_{(j)} = 2 _{(min[2j,N-2j])} + 1 + 1^{'}
\end{equation}

\noindent
(if $j \neq m/2$ )

\begin{equation}
2_{(j)} \times 2_{(m - j)} = 2_{|m - 2j|} + 1^{''} + 1^{'''}
\end{equation}

\noindent
(if $j \neq m/2 $)

\begin{equation}
2_{m/2} \times 2_{m/2} = 1 + 1^{'} + 1^{''} + 1^{'''}
\end{equation}

\noindent
This last is possible only if m is even and hence if N is divisible by {\it
four}.\\

For $Q_{2N}$, there are four singlets $1$,$1^{'}$,$1^{''}$,$1^{'''}$ and
$(N - 1)$ doublets $2_{(j)}$ ($ 1 \leq j \leq (N-1) $).

The singlets have the multiplication rules:

\begin{equation}
1 \times 1 = 1^{'} \times 1^{'} = 1
\end{equation}
\begin{equation}
1^{''} \times 1^{''} = 1^{'''} \times 1^{'''} = 1^{'}
\end{equation}
\begin{equation}
 1^{'} \times 1^{''} = 1^{'''} ; 1^{'''} \times 1^{'} = 1^{''}
\end{equation}

\noindent
for $N = (2k + 1)$ but are identical to those for $D_N$ when N = 2k.

The products involving the $2_{(j)}$ are identical to those given
for $D_N$ (N even) above.

This completes the multiplication rules for 19 of the 45 groups. When needed,
rules for the other groups will be derived.

\newpage

{\bf III Gauged Finite Groups and Anomalies}\\

The models we shall consider have a symmetry comprised of the standard model
gauge group
$SU(3)_C \times SU(2)_L \times U(1)_Y $ producted with a nonabelian finite
group G.

If G is a global (ungauged) symmetry, there are problems if the spacetime
manifold is
topologically nontrivial since it has been shown that any such global
symmetry is broken in the presence of wormholes\cite{global}. From a Local
viewpoint (Local with a capital
means within a flat spacetime neighbourhood) the distinction between a global
and local (gauged) finite symmetry does not exist. The distinction exists only
in a
Global sense (Global meaning pertaining to topological aspects of the
manifold).
In a flat spacetime, gauging a finite group has no meaning. In the
presence of wormholes, themselves expected to be inevitable from the
fluctuations
occurring in quantum gravity, gauging G is essential. The mathematical
treatment of such a gauged finite group has a long history\cite{flat}.

In order to gauge the finite group G, the simplest procedure is to gauge
a continuous group H which contains G as a subgroup, and then to spontaneously
break H by choice of a Higgs potential. The symmetry breaking may occur at a
high
energy scale, and then the low energy effective theory will not contain any
gauge potentials or gauge bosons; this effective theory is, as explained above,
Locally identical
to a globally-invariant theory with symmetry G.

For example, consider G = $Q_6$ and H = $SU(2)$. We would
like to use only one irreducible representation $\Phi$ of $Q_6$
in the symmetry-breaking potential $V(\Phi)$. The irreps. of $Q_6$
are $1, 1', 1^{''}, 1^{'''}, 2, 2_S$. The $1^{''}, 1^{'''}$ and
$2_S$ are spinorial and appear in the decompositions only of
$2, 4, 6, 8 ....$ of $SU(2)$. Since $\Phi$ must contain the $1$
of $Q_6$ we must choose from the vectorial irreps. $3, 5, 7, 9 ...$
of $SU(2)$. The appropriate choice is the $7$ represented by a symmetric
traceless
third-rank tensor $\Phi_{ijk}$ with $\Phi_{ikk} = 0$.

For the vacuum expectation value, we choose
\begin{equation}
<\Phi_{111}> = +1; <\Phi_{122}> = -1
\end{equation}
and all other unrelated components vanishing. If we look for the $3\times3$
matrices $R_{ij}$ whicg
leave invariant this VEV we find from choices of indices in
\begin{equation}
R_{il}R_{jm}R_{kn}<\Phi_{lmn}> = <\Phi_{ijk}>
\label{eq18}
\end{equation}
that $R_{31} = R_{32} = 0$ (Use $<\Phi_{3ij}\Phi_{3ij}> = 0$) and that $R_{33}
= \pm1$. Then we find $(R_{11})^3 - 3R_{11}(R_{12})^2 = 1$ (Use $l = m = n = 1$
in $(\ref{eq18})$). This means that if $R_{11} = \cos\theta$ then $\cos3\theta
= 1$ or $\theta_n = 2\pi n/3$. So the elements of $Q_6$ are $A = R_3(\theta_1),
A^2, A^3$ and $B, BA, BA^2$ where $B =$ diag$(i, -i -i)$.

More generally, it can be shown that to obtain $Q_{2N}$ one must use an $N^{\rm
th}$ rank
tensor because one finds for the elements $R_{11}$ and $R_{12}$:
\begin{equation}
\sum_{p=0}^{[N/2]}(-1)^p{N\choose2p} (R_{11})^{N-2p}(R_{12})^{2p} =
\cos N\theta = 1
\end{equation}

If the group H is gauged, it must be free from anomalies. This entails several
conditions which must be met:

(a) The chiral fermions must fall into complete irreducible representations
not only of G but also of H.

(b) These representations must be free of all H anomalies including $(H)^3$,
$(H)^2Y$;
for the cases of H = $O(3), SU(2)$ only the latter anomaly is nontrivial.

(c) If H = $SU(2)$, there must be no global anomaly.

The above three conditions apply to nonabelian H. The case of an abelian H
avoids (a) and (c) but gives rise to additional mixed anomalies in (b).

For nonabelian H, conditions (b) and (c) are straightforward to write down and
solve.
Condition (a) needs more discussion. We shall focus on the special cases
of $O(3) \supset D_N$
and $SU(2) \supset Q_{2N}$.

For $O(3)$ the irreps. are ${\bf 1,3,5,7,....}$ dimensional. $D_N$ has irreps.
(for even $N = 2m$) $1, 1^{'}, 1^{''}, 1^{'''}$ and $2_{(j)} (1 \leq j \leq (m
- 1))$ and these correspond
to:
\begin{equation}
O(3):  {\bf 1}  \rightarrow 1 ; {\bf 3} \rightarrow (1^{'} + 2_{(1)})
\end{equation}

\noindent

The same situation occurs for odd N with irreps. $1, 1^{'}$ and $2_{(j)} ( 1
\leq j \leq (N - 1)/2)$.
If we insist on keeping within the fermions of the standard model, or as close
to that
ideal as possible, nothing beyond a {\bf 3} is necessary because the same
quantum
numbers are not repeated more than three times.

For $SU(2) \supset Q_{2N}$ the corresponding breakdown is:
\begin{equation}
{\bf 1} \rightarrow 1; {\bf 2} \rightarrow 2_{S(1)} ; {\bf 3} \rightarrow 1^{'}
+ 2_{(1)}
\end{equation}
\noindent
where the doublets of $Q_{2N}$, $2_{(1)}$ and $2_{S(1)}$, are defined by Eq.
(20).

These are the principal splittings of a continuous group irrep. into finite
subgroup irreps. we shall
need in our discussions of model building below.

\newpage

{\bf IV Model Building in General}\\

In order to be specific we need to set up a collection of model-building rules.
The main
purpose is to understand why the third family of quarks and leptons is heavy,
and especially why the top
quark is {\bf very} heavy. Thus we require that:

(A) The t quark mass (and {\it only } the t ) transforms as a {\bf 1} of G.

(B) The b and $\tau$ masses appear as G is broken to $G^{'}$.

We next require that at tree level or one-loop level the second family be
distinguishable from the
first. That is:

(C) After stage (B) first the c mass ($G' \rightarrow G''$), then the
s and $\mu$ masses ($G^{''} \rightarrow
G^{'''}$) are generated.

At stage (C) the u, d and e remain massless.

Inaddition to the above constraints we require that:

(D) No additional quarks and a minimal number of leptons be introduced beyond
the usual three-family
standard model.

(E) All anomalies are cancelled as described in Section III above, when G is
embedded in the minimal
continuous Lie group H:  $G \subset H$.

We strive to satisfy all of (A) through (E). By the study of specific cases in
the following Section V
we shall see that these constraints are quite nontrivial to satisfy
simultaneously and that
the number of interesting models is small (We shall arrive at only {\it two}).

\newpage

{\bf V Specific Examples}\\

We shall treat special cases for the  nonabelian\cite{ross} group G in turn,
taken from the complete
listing, up to order g = 31, given in Section II above.\\

(a) {\it Dihedral Groups ($D_N$, order g = 2N)}\\

{}From the multiplication rules for $D_N$ we see that if the top quark mass
transforms as a singlet,
as required by rule (A) above, the $t_L$ and $t_R$ must {\it both} be in $1$ or
$1^{'}$ or the
{\it same} $2_{(j)}$. The doublet is unsuitable because it will include a
second quark, violating
rule (A).

To proceed systematically, note that there are three triples of quarks with
common
quantum numbers: (1) $(t,b)_L,(c,s)_L,(u,d)_L$; (2) $t_R,c_R,u_R$ ; and (3)
$b_R,s_R,d_R$.
Since $D_N$ is a subgroup of $SO(3)$ we must look to the rule (E) to see that
each triple must
be in $1+1+1$ or $1^{'}+2_{(1)}$ (equivalent to the vector {\bf 3} of $O(3)$ as
can be deduced
from the $O(3)$ and $D_N$ character tables) of $D_N$ to avoid anomalies. If
$t_L$ and $t_R$
are in $1^{'}$ it follows that $(c,s)_L$, $(u,d)_L$ and $c_R,u_R$ are in
$2_{(1)}$ implying that
charm and up quarks have singlet components in their mass terms, hence
violating rule (A).
If $t_L$ and $t_R$ are in $1$, then so are $(c.s)_L,(u,d)_L$ and $c_R, u_R$
again violating rule (A).

{}From this discussion we deduce that no suitable model based on G = $D_N$
exists.\\

(b) {\it Permutation Groups ($S_N$, order g = N!)}\\

The group $S_3$ is identical to $D_3$ which was excluded in (a) above.
The only other $S_N$ with $g \leq 31$ is $S_4$ which has irreducible
representations
$1, 1^{'}, 2, 3, 3^{'}$. It is a subgroup of $O(3)$ so the triples must be in
$1+1+1$
or $3$ ($5$ of $O(3) \rightarrow 2 + 3^{'}$). Since $3 \times 3 = 1 + 2 + 3 +
3^{'}$ contains
a singlet, neither choice fulfils rule (A).

Hence the groups G = $S_N$ are excluded.\\

(c) {\it Tetrahedral Group (T, order g = 12)}\\

The group T has $1, 1^{'}, 1^{''}, 3$ representations
and $T \subset O(3)$ with irreps. of $O(3)$
decomposing under T as $1 \rightarrow 1$, $3 \rightarrow 3$,
$5 \rightarrow 1^{'} + 1^{''} + 3$.
Thus $t_L$ and $t_R$ must either both be in 1 or 3 and since
$3 \times 3 = 1 + 1^{'} + 1^{''} +2(3)$
both choices violate rule (A).

Hence T cannot be used.\\

(d) {\it Double Dihedral (or Dicyclic) Groups ($Q_{2N}$, order g = 4N)}\\

The above cases (a),(b),(c) are all subgroups of $O(3)$. There exist
counterparts
which are doubled and are subgroups {\it not} of $O(3)$ but of $SU(2)$.

As a first example, consider ${ }^{(d)}D_N = Q_{2N}$ which has representations
$1, 1^{'}, 1^{''}, 1^{'''}$ and $(N - 1)$ doublets $2_{(j)}$
as described above in Section II. In $SU(2)$
the representations decompose under $Q_{2N}$ as: $1 \rightarrow 1, 2
\rightarrow 2_{(2)}$
and $3 \rightarrow 1^{'} + 2_{(1)}$. Thus the possible choices for the three
quark triples are: $1 + 1 + 1$,
$1 + 2_{(2)}$ and $1^{'} + 2_{(1)}$.

We can now go a long way toward fulfilling all the rules in Section IV above.

In the quark sector $t_L$ and $t_R$ must be both $1$ or both $1^{'}$. The
latter leads to other singlet
mass terms from $2_{(1)} \times 2_{(1)}$ and so is excluded. Thus $t_L$ and
$t_R$ must both be $1$.
The simplest choice is then to use $Q_6$ ($Q_{2N}, N \geq 4$ leads to no new
structure) and then there
are two different assignments within $Q_6$, namely:\\

Choice A.\\

$$\begin{array}{ccccccc}

\left( \begin{array}{c} t \\ b \end{array} \right)_{L} &
1 & \begin{array}{c} t_{R}~~~ 1 \\ b_{R} \hspace{0.2in}1^{'} \end{array} &
\left( \begin{array}{c} \nu_{\tau} \\ \tau \end{array} \right)_{L} & 1 &
\tau_{R} & 1^{'} \\

\left. \begin{array}{c} \left( \begin{array}{c} c \\ s \end{array} \right)_{L}
\\
\left( \begin{array}{c} u \\ d \end{array} \right)_{L} \end{array}  \right\} &
2_{S}
 &  \begin{array}{c} \left. \begin{array}{c} c_{R} ~~~ 1\\ u_{R} ~~~ 1
\end{array} \right. \\
\left. \begin{array}{c} s_{R} \\ d_{R} \end{array} \right\} 2 \end{array} &
\left. \begin{array}{c} \left( \begin{array}{c} \nu_{\mu} \\ \mu \end{array}
\right)_{L} \\
\left( \begin{array}{c} \nu_{e} \\ e \end{array} \right)_{L} \end{array}
\right\} & 2_{S}
& \left. \begin{array}{c} \mu_{R} \\ \\ e_{R} \end{array} \right\} & 2

\end{array}$$\\

Choice B.

$$\begin{array}{ccccccc}

\left( \begin{array}{c} t \\ b \end{array} \right)_{L} &
1 & \begin{array}{c} t_{R}~~~ 1 \\ b_{R} \hspace{0.2in}1^{'} \end{array} &
\left( \begin{array}{c} \nu_{\tau} \\ \tau \end{array} \right)_{L} & 1 &
\tau_{R} & 1^{'} \\

\left. \begin{array}{c} \left( \begin{array}{c} c \\ s \end{array} \right)_{L}
{}~~~ 1 \\
\left( \begin{array}{c} u \\ d \end{array} \right)_{L} ~~~ 1 \end{array}
\right.  &
 &  \begin{array}{c} \left. \begin{array}{c} c_{R} \\ u_{R} \end{array}
\right\} 2_{S} \\
\left. \begin{array}{c} s_{R} \\ d_{R} \end{array} \right\} 2 \end{array} &
\left. \begin{array}{c} \left( \begin{array}{c} \nu_{\mu} \\ \mu \end{array}
\right)_{L} ~~~ 1 \\
\left( \begin{array}{c} \nu_{e} \\ e \end{array} \right)_{L} ~~~ 1 \end{array}
\right.&
& \left. \begin{array}{c} \mu_{R} \\ \\ e_{R} \end{array} \right\} & 2

\end{array}$$\\

Both choices A and B are free from $(SU(2)^{'})^3$ anomalies. Now consider the
mixed $(SU(2)^{'})^2Y$
anomaly. It is nonvanishing for both A and B. This is inevitable for any
assignment even for
the case of $Q_6$ as can be seen by studying the following table. Normalize the
quadratic Casimir
of the $SU(2)^{'}$ doublet to $+1$ and hence the triplet to $+4$; define
$Q = T_3 + Y$ and the anomaly is:\\

$$\begin{tabular}{||c||c|c|c||}   \hline
 &$2_{S} + 1$ &$ 3$ & $1 + 1 + 1$ \\    \hline\hline
$\left( \begin{array}{c} \nu \\ e^{-} \end{array} \right)_{Li}$ &$ -1$ &$ -4$
&$ 0$   \\
$e^{+}_{Li}$ & $+1$& $+4$ & $0$ \\    \hline
$\left( \begin{array}{c} u \\ d \end{array} \right)_{Li} $& $+1$ & $+4$ & $0$
\\
$\bar{u}_{Li}$ &  $-2$ & $-8$ & $0$ \\
$\bar{d}_{Li}$ & $+1$ &  $+4$ & $0$   \\  \hline
\end{tabular}$$\\

Thus this anomaly adds to $+8$ for Choice A, $+6$ for Choice B. To cancel this
n the most economial waye, can add appropriate additional leptons. From the
above Table, we see that
for all choices of $Q_6$ representations we need to extend slightly the lepton
sector whereupon
cancellation of all anomalies is always possible.

For example, in choice A the anomaly $+8$ can be compensated by adding two {\bf
3}s of $SU(2)^{'}$
of left-handed leptons and corresponding singlets of right-handed leptons with
usual quantum
numbers. The quark sector is exactly as in the standard model. The additional
particles might be called Q-leptons. Because their masses break $SU(2)_L$,
their masses should be below about $200$GeV; but phenomenology
dictates that they be above $50$GeV.

For the Choice A the mass matrices are:\\

$$U = \left( \begin{tabular}{c|c}
$<2_S>$ & $ <2_S> $ \\  \hline
$<1>  $ & $ <1>   $
\end{tabular} \right)$$ \\

and:\\

$$D = L = \left( \begin{tabular}{c|c}
$<1''+1'''+2_S>$ & $ <2_S> $ \\  \hline
$<2>  $ & $ <1'>   $
\end{tabular} \right)$$\\
The way of implementing the hierarchy is by the following steps which comply
with
the rules (A) - (D) of Section IV above:

(A) A VEV to an $SU(2)_L$ doublet which is a singlet of $Q_6$ gives t its heavy
mass without breaking
$Q_6$.

(B) A VEV to a $1^{'}$ gives b and $\tau$ their masses, at the same time
breaking $G = Q_6$ to
$G^{'} = Z_6$.

(C) and (D) The charm quark mass is radiatively generated according to the
diagram of Fig. (1)
which uses a VEV transforming as $(1,2_{S})$ under $SU(2)_L \times SU(2)^{'}$.
Next
the s and $\mu$ acquire their tree-level masses through a $(2, 1^{''} or
1^{'''})$ VEV. These VEVs
break $Z_6$ completely. At this point the u,d and e are still massless.\\

(e) {\it Double Tetrahedral Group (${~~}^{(d)}T$, order g = 24)}\\

The doubled group ${ }^{(d)}T \subset SU(2)^{'}$ has representations $1, 1^{'},
1^{''}, 2_{S}, 2^{'}_{S}, 2^{''}_{S}$ and $3$.
The lowest dimensional representations of $SU(2)^{'}$ decompose as:
$1 \rightarrow 1$,$2 \rightarrow 2_{S}$, $3 \rightarrow 3$.
This leads to a model quite analogous to the $Q_6$ model described above with
the same advantages. Note
that ${ }^{(d)}T$ is isomorphic to $Z_3 \tilde{\times} Q_4$. We assign:

$$\begin{array}{cc}

\left. \begin{array}{c} \left( \begin{array}{c} t \\ b \end{array} \right)_{L}
{}~~~ 1\\
\left. \begin{array}{c} \left( \begin{array}{c} c \\ s \end{array} \right)_{L}
\\
\left( \begin{array}{c} u \\ d \end{array} \right)_{L}  \end{array} \right\}
2_{S} \end{array} \right.&
\left. \begin{array}{c} t_{R}~~~ 1 \\ c_{R} ~~~ 1 \\ u_{R} ~~~ 1 \\
\left.\begin{array}{c}
 b_{R} \\ s_{R} \\ d_{R} \end{array} \right\} 3 \end{array} \right.
\end{array}$$  \\

$$\begin{array}{cc}

\left. \left( \begin{array}{c} \nu_{\tau} \\ \tau \end{array} \right)_{L} ~~~ 1
\right.\\
\left.  \begin{array}{c} \left( \begin{array}{c} \nu_{\mu} \\ \mu \end{array}
\right)_{L}  \\
\left( \begin{array}{c} \nu_{e} \\ e \end{array} \right)_{L} ~~~  \end{array}
\right\}  2_{S}  &
\left. \begin{array}{c} \tau_{R} \\ \mu_{R} \\ e_{R} \end{array} \right\} 3
\end{array} $$   \\

whereupon the mass matrices are:

$$U = \left( \begin{tabular}{c|c}
$<2_S>$ & $ <1> $ \\  \hline
$<2_S>  $ & $ <1>   $
\end{tabular} \right)$$ \\

and:\\

$$D = L = \left( \begin{tabular}{c|c}
$<2_S + 2^{'}_S + 2^{''}_S>$ & $ <3> $ \\  \hline
$<2_S + 2^{'}_S + 2^{''}_S>$ & $ <3> $
\end{tabular} \right)$$\\

To implement the hierarchy complying with rules (A) to (D) of Section IV gives:

(A) A VEV to a $SU(2)_L$ doublet which is a singlet of ${ }^{(d)}T$ gives a
heavy mass to t
without breaking ${ }^{(d)}T$.

(B) A VEV to a {\bf 3} of ${ }^{(d)}T$ gives mass to b and $\tau$.

(C) and (D)  The c quark acquires mass radiatively through
a VEV of $(1 , 2_S )$ via the diagram of Fig. (1). The s and $\mu$ acquire mass
at tree level
through $2^{'}_S$ or $2^{''}_S$ VEVs, breaking $G^{'}$. The u, d and e are
still massless.\\

(f) {\it Other Groups}\\

We have so far fealt with 22 of the 45 nonabelian groups with $g \leq 31$.
Another 11 are not
simple so fall outside our search. The remaining 12 are twisted products of
$Z_N$'s and of
orders: g = 16 (5), 18, 20, 21, 24 (2) and 27 (2). Note that one of these has
already been
discussed at length since ${ }^{(d)}T = Z_3 \tilde{\times} Q_4$.

Of the remainder, $Z_9 \tilde{\times} Z_3 = \Delta(27)$, a subgroup of $SU(3)$.
None of the others is embeddable in an $SU(2)^{'}$, since all such groups are
considered in (a) -(e)
above. The only other one which embeds in $SU(3)$ is $\Delta(24) = T \times
Z_2$
but triplets do not allow a hierarchy following our rules of Section IV. The
only other Lie
group of interest with irreducible representations $\leq 3$ is $SO(4) = SU(2)
\times SU(2)$
with a simple direct product. None of our list embeds minimally in $SO(4)$
because their
products are twisted, and not simple.

Thus only two possibilities - $Q_{2N}$ ($N \geq 3$) and ${ }^{(d)}T$
- permit a fermion hierarchy of the type we have specified.\\

\newpage

{\bf VI Concluding Remarks}\\

We have exhibited models with a nonabelian discrete flavor group G where G can
be embedded in an anomaly-free
$SU(2)$. The top quark can be much heavier than all other quark flavors because
it alone has a G-invariant mass.

The breaking of G gives rise sequentially to the other fermion masses: first b
and $\tau$; then c; and finally
s and $\mu$. The first family masses are so tiny that they are neglected at the
order considered here. For
the simplest case of $G = Q_6$, a testable prediction is the occurrence of the
Q-leptons
in the mass range $50$GeV to $200$GeV.

We should mention the point that the spontaneous breaking of discrete
symmetries
always gives rise to the danger of unacceptable cosmological
domain walls\cite{zel}. However this danger can always be avoided by adding
explicit soft breaking in the potential function.

Many outstanding questions remain such as: Can this scheme be made more
quantitative?
How would supersymmetry effect the model building? How does G arise in a more
complete framework\cite{SU3}?

Although we have removed the extreme hierarchy of the Yukawa couplings, it has
been replaced by a
more involved Higgs sector. This seems inevitable; the point is that the
discrete group G leads
to a new viewpoint that provides a first step to understanding the mass
spectrum of quarks and leptons,
particularly why the top quark mass is so different from all other fermion
masses.\\

One of us (T.W.K.) thanks the members of the Institute of Field Physics
at UNC-Chapel Hill for their generous hospitality while this work was
in progress.  This work was supported in part by the U.S. Department of
Energy under Grants DE-FG05-85ER-40219 and DE-FG05-85ER-40226.

\newpage

{\bf Figure Caption}\\

Fig. 1   One loop diagram contributing to the charm quark mass.

\newpage

\begin{flushleft}
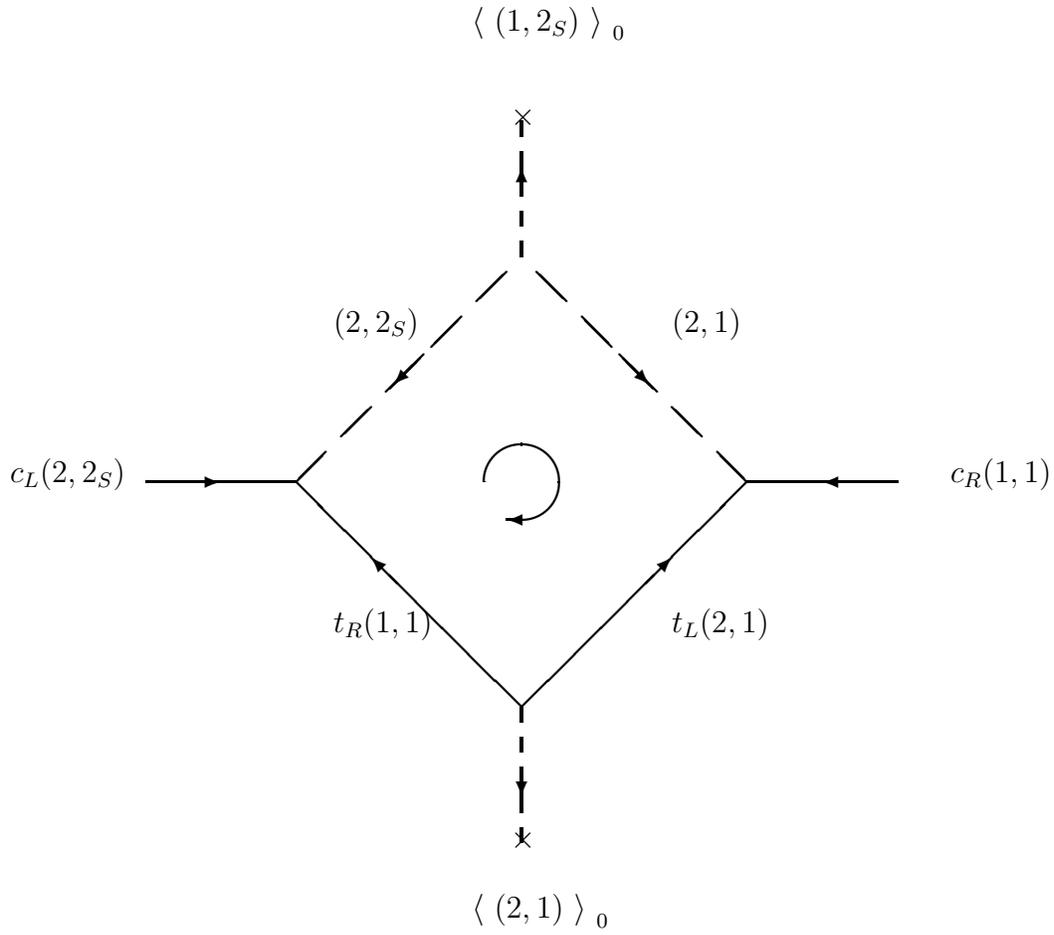
\begin{figure}[h]

\vspace*{2.0cm}

\setlength{\unitlength}{1.0cm}

\begin{picture}(15,12)

\thicklines

\put(2,6){\thicklines\vector(1,0){1}}
\put(3,6){\line(1,0){1}}

\multiput(4,6)(0.6,0.6){5}{\line(1,1){0.4}}

\put(7,3){\vector(-1,1){2}}
\put(5,5){\line(-1,1){1}}

\multiput(7,9)(0,0.4){5}{\line(0,1){0.2}}

\put(12,6){\vector(-1,0){1}}
\put(11,6){\line(-1,0){1}}

\multiput(10,6)(-0.6,0.6){5}{\line(-1,1){0.4}}

\multiput(7,3)(0,-0.4){5}{\line(0,-1){0.2}}

\put(7,3){\vector(1,1){2}}
\put(9,5){\line(1,1){1}}

\put(8.3,7.7){\vector(1,-1){0.4}}

\put(5.7,7.7){\vector(-1,-1){0.4}}

\put(7,9.8){\vector(0,1){0.4}}

\put(7,2.2){\vector(0,-1){0.4}}

\put(6.854,1.127){${\bf \times}$}

\put(6.854,10.75){$\times$}

\put(6.2,12){
$ \left \langle \ {(1,2_S)} \ \right \rangle_{\ 0} $}

\put(6.2,0.2){
$ \left \langle \ {(2,1)} \ \right \rangle_{\ 0} $}

\put(0.2,6){$ c_L (2,2_S) $}

\put(12.7,6){$ {c_R (1,1)} $}

\put(7,6){\oval(1,1)[t]}
\put(7,6){\oval(1,1)[br]}
\put(7,5.5){\vector(-1,0){0.2}}

\put(9,8){$(2,1)  $}
\put(9,4){$ t_L (2,1) $}

\put(4.5,8){$ (2,2_S) $}
\put(4.5,4){$ t_R (1,1)$}

\end{picture}

\caption{One loop diagram contributing to the charm quark mass.}

\label{Fig. 1}

\end{figure}

\end{flushleft}


\begin{references}
\bibitem{guts}
Another (non-minimal by the above definition) route to discrete family
symmetry is through the previous $SU(N)$ family symmetry models\cite{SUN}
. Here one can consider the breaking pattern
\begin{equation}
SU(N)\rightarrow SU(5)\times SU(5-N)\times U(1)\rightarrow SU(5)\times G
\end{equation}
where G is a discrete group. Such models typically have more than just
three complete families until $G$ is broken. Although worthy of further
study, these models do not in general satisfy our minimality condition
on the fermion spectrum.
\bibitem{rootN}
Note that even though N 3-vector VEVs are needed to break $SO(3)$ to $D_N$
the top Yukawa has a factor $N^{-1/2}$ suppression as can be seen by
comparing the t and W masses.
\end{references}
\end{document}